# Far-from-equilibrium processes without net thermal exchange via energy sorting

Jose M. G. Vilar[1,2,*] and J. Miguel Rubi[3]

[1]Biophysics Unit (CSIC-UPV/EHU) and Department of Biochemistry and Molecular Biology, University of the Basque Country, 48080 Bilbao, Spain

[2]IKERBASQUE, Basque Foundation for Science, 48011 Bilbao, Spain

[3]Departament de Fisica Fonamental, Universitat de Barcelona, Diagonal 647, 08028 Barcelona, Spain

[*]Correspondence: j.vilar@ikerbasque.org

## Abstract

**Many important processes at the microscale require far-from-equilibrium conditions to occur, as in the functioning of mesoscopic bioreactors, nanoscopic rotors, and nanoscale mass conveyors. Achieving such conditions, however, is typically based on energy inputs that strongly affect the thermal properties of the environment and the controllability of the system itself. Here, we present a general class of far-from-equilibrium processes that suppress the net thermal exchange with the environment by maintaining the Maxwell-Boltzmann velocity distribution intact. This new phenomenon, referred to as ghost equilibrium, results from the statistical cancellation of superheated and subcooled nonequilibrated degrees of freedom that are autonomously generated through a microscale energy sorting process. We provide general conditions to observe this phenomenon and study its implications for manipulating energy at the microscale. The results are applied explicitly to two mechanistically different cases, an ensemble of rotational dipoles and a gas of trapped particles, which encompass a great variety of common situations involving both rotational and translational degrees of freedom.**

Effective manipulation of energy at the microscale is fundamental to create the far-from-equilibrium conditions needed for the occurrence of many micron and submicron processes. Transferring energy into small-scale systems is prone to affect the thermal properties of the environment and the controllability of the system itself. The result is often very large thermal gradients, which typically range from $10^5$ K/m on the mesoscale (1) to $10^8$ K/m on the nanoscale (2). The generation of thermal gradients and temperature differences arises in the first steps toward establishing the fluxes and forces that ultimately lead to motion, currents, and many other processes that can perform useful tasks at the microscale. Thermal gradients, have been used, for instance, to drive subnanometer motion of cargoes along carbon nanotubes (3) and are present in nanoscale mass conveyors (4). Down to the molecular level, thermal gradients can polarize water molecules and obtain large electric fields (5). Thermal heterogeneities have also been used to trap DNA in mesoscopic bioreactors (6).

Motivated by the relevance of energy manipulation in such wide-ranging processes, we asked whether it is possible to drive systems far from equilibrium with strong internal temperature differences but without net thermal exchange with their environment. We have focused on the most robust case in which the overall thermal energy distribution for the far-from-equilibrium system is shaped as the equilibrium Maxwell-Boltzmann velocity distribution. This type of approach will provide the system as a whole with the same kinetic energy distribution as the thermal bath, thus minimizing the perturbation of the environment. The Maxwell-Boltzmann velocity distribution (7) plays a central role because it is a universal feature of systems at equilibrium (8), it determines key thermodynamic properties, and its absence is a signature of far-from-equilibrium conditions (9, 10). Therefore, it has tacitly been accepted that its presence implies close proximity to equilibrium (11).

Here, we show that there exists a new type of states that are deep inside the nonequilibrium regime but that, paradoxically, maintain the Maxwell-Boltzmann velocity distribution and global temperature as if they were exactly at equilibrium. This new phenomenon, which we have termed ghost equilibrium, results from the self-sorting of the kinetic energy into superheated and subcooled regions along a nonequilibrated degree of freedom that statistically cancel each other for the whole system.

We consider the general class of systems described by a coordinate, $s$, and its associated velocity, $u$, with dynamics governed by the stochastic second-order differential equation

$$\frac{ds}{dt} = u,$$
$$\frac{du}{dt} = -\gamma u + F(s,t) + \Gamma(t), \tag{1}$$

where $F(s,t)$ is a force-dependent function of the coordinate and time, and $\gamma$ is a damping constant. Fluctuations are taken into account by a Gaussian noise term $\Gamma(t)$ with zero mean and correlation function $\langle \Gamma(t)\Gamma(t') \rangle = 2D\delta(t-t')$. The intensity of the noise, $D$, is determined by the thermal energy $kT$ at equilibrium by the fluctuation-dissipation theorem: $D = \gamma kT / \mu$, where $\mu$ is a parameter quantifying the inertia of the degree of freedom. This stochastic equation, known as Langevin equation, captures the fluctuating behavior of a wide diversity of dynamical systems in contact with a thermal bath. It was originally proposed for Brownian motion (12) and it has subsequently been extended to many situations (13), including electronic devices (14, 15), organic semiconductors (16), confined colloids (17), suspensions of ferromagnetic particles (18), granular matter (19), nuclear physics (20), critical dynamics (21), and biological systems (22-25).

Close to equilibrium, the velocity distribution of systems governed by Equation (1) is always Gaussian for each value of $s$ (9). It is obtained from the probability density of $s$ and $u$, which for sufficiently slow changes of $F(s,t)$ is given by

$P(s,u) = e^{-\gamma \frac{U(s,t)+u^2/2}{D}} / \int_{-\infty}^{\infty} e^{-\gamma \frac{U(s,t)+u^2/2}{D}} ds\, du$, where $U(s,t) = -\int_{s_0}^{s} F(s',t) ds'$ can be viewed as a generalized potential energy. In general, the velocity distribution is not Gaussian and the exact form of the probability distribution is determined by the Fokker-Planck equation (26) $\partial_t P(s,u) = \left[ -\partial_s u + \partial_u \left( \gamma u - F(s,t) \right) + D\partial_{u,u} \right] P(s,u)$.

## Results

**General conditions.** To concentrate on the velocity distribution, we coarse-grain the probability density between two values, $s_1$ and $s_2$, of the degree of freedom $s$:

$$\tilde{P}(u) = \frac{\int_{s_1}^{s_2} P(s,u)ds}{\int_{-\infty}^{\infty} du \int_{s_1}^{s_2} P(s,u)ds}.$$  (2)

The dynamics of this reduced probability, after substitution in the Fokker-Planck equation, is given by

$$\partial_t \tilde{P}(u) = \left( \partial_u \gamma u + D \partial_{u,u} \right) \tilde{P}(u)$$
$$- \partial_u \int_{s_1}^{s_2} F(s,t) P(s,u)ds - u(P(s_2,u) - P(s_1,u)).$$  (3)

This non-closed equation shows that, in general, the velocity distribution evolves with time and would not keep its Gaussian shape far from equilibrium. If the right-hand side of the equation is identically zero, however, the velocity distribution in the coordinate-space region between $s_1$ and $s_2$ would remain constant. We focus on conditions that cancel out each of its three terms simultaneously.

The drift-diffusion term, $\partial_u (\gamma u \tilde{P}(u)) + D \partial_{u,u} \tilde{P}(u)$, is zero for Gaussian velocity distributions of the type $\tilde{P}(u) = e^{-\gamma u^2/(2D)} \big/ \int_{-\infty}^{\infty} e^{-\gamma u^2/(2D)} du$, which are present for systems at equilibrium at time $t_0$. The force-dependent term, $\partial_u \int_{s_1}^{s_2} F(s,t) P(s,u)ds$, vanishes when force is switched off at time $t_0$ so that $F(s,t) = F(s)\Theta(t - t_0)$, where $\Theta$ is the Heaviside step function.

The boundary term, $P(s_2,u) - P(s_1,u)$, cancels out in two important situations. The first one is when the system has periodicity $\Lambda$ in the degree of freedom so that $s_2$ is equivalent to $s_1 + \Lambda$. This periodicity straightforwardly guarantees the cancellation of this term. The second one is when the system has specular symmetry and reflecting boundary conditions. Specular symmetry around $s_0$ implies $P(s_0 - s, u) = P(s_0 + s, -u)$, which can be used to rewrite the boundary term, with $s_1 = s_0 - L$ and $s_2 = s_0 + L$, as $P(s_0 + L, u) - P(s_0 + L, -u)$, which in turn cancels out for reflecting boundary conditions.

Both a periodic system and a specular system with reflecting boundaries initially at equilibrium will cancel out the right hand side of Equation (3) term by term upon switching off the effects of the force (Fig. 1). Therefore, they will maintain their Maxwell-Boltzmann velocity distribution intact during the relaxation to the new equilibrium state. If the inertial effects are relevant, the relaxation process will take place beyond local equilibrium (10).

**Application to rotational dipoles.** As an exemplar of periodic system (Fig. 1a), we consider the rotational motion of an ensemble of non-interacting dipoles in an external field (9, 27). This type of systems encompasses small colloidal ferromagnetic particles, electric dipoles, and microorganisms that act as gravitational dipoles. In this case, $s = \theta$ is the angle between the dipole and the field, $u = \omega$ is the angular velocity, $\mu = I$ is the moment of inertia, and $\gamma$ is the rotational damping. The force-dependent function is given by $F(\theta) = -pE \sin(\theta) / I$, where $p$ is the dipolar moment and $E$ is the field.

As predicted by our general analysis, the system maintains the Maxwell-Boltzmann angular velocity distribution intact upon switching off the field (Fig. 2a). The system, however, evolves deep inside the nonequilibrium regime. The initial shape of the velocity distribution is not generally kept during the relaxation process for any region of angular coordinate space that does not cancel out the boundary term of the right hand side of Equation (3). Indeed, the velocities redistribute along the angular coordinate. The distribution gets broader for the ensemble of dipoles with $\cos(\theta) < 0$ (Fig. 2b) and thinner for dipoles with $\cos(\theta) > 0$ (Fig. 2c). When the system gets close to the final equilibrium state, the Maxwell-Boltzmann velocity distribution settles in again uniformly for every region of the angular coordinate space.

The redistribution of velocities during the relaxation process leads to an autonomous sorting of the kinetic energy in the angular-coordinate space with regions that transiently heat up or cool down (Fig. 3). This behavior at the microscale, so far away from equilibrium, strongly contrasts with the global behavior, for which the system keeps its kinetic energy and temperature intact as if it were at equilibrium. Thus, the system can be viewed as being in a ghost equilibrium state, which results from the precise statistical cancellation of the effects of the superheated and subcooled regions.

**Application to a gas of trapped particles.** Our general results show that systems with specular symmetry and reflecting boundaries can also be in a ghost equilibrium state. A prototypical such system is a gas of non-interacting particles trapped in a parabolic potential inside a box (Fig. 1b), which can be implemented explicitly by an ensemble of polystyrene beads in the force field of an optical trap (28-30). In this case, $s = x$ corresponds to the position; $u = v$, to the velocity; $\mu = m$, to the mass; and $\gamma$, to the velocity damping. The force-dependent function is given by $F(x) = -Kx / m + \infty(\delta(x + a) - \delta(x - a))$, where $K$ is the force constant of the trap and $a$ indicates the distance of the box walls from the trap center.

Upon switching off the trap, the behavior of this system along the translational coordinate closely parallels that of the ensemble of dipoles along the angular coordinate. The system globally maintains the Maxwell-Boltzmann distribution intact (Fig. 4a) but the distribution gets wider in the region of the translational-coordinate space close to the box walls (Fig. 4b) and thinner next to the center (Fig. 4c). Similarly, there is also sorting of the kinetic energy into different regions so that the system heats up towards the box walls and cools down in the central region whereas the global temperature remains constant during the whole relaxation process (Fig. 5).

In general, if the system does not have the required symmetries, the time evolution of the velocity distribution will not keep the Gaussian form. This type of cases is present when the center of the trap is not in the middle of the box, as for instance when it is next to the box wall (Fig. 6a) with a force term $F(x) = -K(x - a) / m + \infty(\delta(x + a) - \delta(x - a))$. In such a case, upon switching off the trap, the velocity distribution becomes asymmetric before developing a shoulder that eventually disappears when the system approaches the new equilibrium state (Fig. 6b), which clearly shows that the relaxation process takes place far from equilibrium.

## Discussion

A major consideration in establishing the far-from-equilibrium conditions needed to drive many important processes at small scales is the effect of the underlying energy sources. Our results have uncovered general conditions that guarantee zero net thermal exchange with the environment by maintaining the equilibrium Maxwell-Boltzmann velocity distribution intact for the overall system, irrespective of strong internal temperature heterogeneities.

These conditions applied explicitly to two systems as pervasive as an ensemble of dipoles and a gas of trapped particles, which involve diverse types of symmetries and degrees of freedom, reveal that the presence of universal equilibrium properties does not necessarily imply that the system is very close to equilibrium. In both cases, there exists a new type of states that are deep inside the nonequilibrium regime but that, paradoxically, look as if they were exactly at equilibrium from the global kinetic energy point of view. The self-sorting of the kinetic energy along a coordinate space in a way that preserves the overall Maxwell-Boltzmann velocity distribution guarantees that there is no net thermal exchange with the environment. These findings, thus, provide new scenarios for manipulating energy at the microscale and open up the possibility of harnessing the effects of temperature differences along the internal structure of the system (3) without changes in its global temperature.

## Acknowledgements

Support was provided by the MICINN under grants FIS2009-10352 (J.M.G.V.) and FIS2008-04386 (J.M.R).

## Author contributions

J.M.G.V. and J.M.R. performed the research and wrote the paper.

*applications* (Springer-Verlag, Berlin).

## Figure captions

**Figure 1. Far-from-equilibrium processes with thermal equilibrium-like distributions can be generated via energy sorting from time-dependent potential energies.** The potential energies of (**a**) a periodic system and (**b**) a specular system with reflecting boundaries are shown before (left) and after (right) switching off the force field. Systems with these types of potential energy functions that are initially in equilibrium will maintain their Maxwell-Boltzmann velocity distributions intact during the relaxation to the new equilibrium states.

**Figure 2. Autonomous sorting of the equilibrium rotational kinetic energy into nonequilibrated degrees of freedom upon switching off the field for an ensemble of dipoles.** In terms of the dimensionless time, $\tau = t \big/ \sqrt{I / kT}$, and angular velocity, $\varpi = \omega \big/ \sqrt{kT / I}$, the dynamics is given by $d\theta / d\tau = \varpi$ and

$d\varpi / d\tau = -\gamma\sqrt{I / kT}\,\varpi - (pE / kT)\Theta(\tau)\sin\theta + \Gamma(\tau)$ with

$\left\langle \Gamma(\tau)\Gamma(\tau') \right\rangle = 2\gamma\sqrt{I / kT}\delta(\tau - \tau')$. The angular velocity distributions are shown at different times after switching off the field for (**a**) the whole system, $\theta \in \left(0, 2\pi\right)$, and for the regions of the angular coordinate with (**b**) $\cos(\theta) < 0$ and (**c**) $\cos(\theta) > 0$. The surfaces on the top-left of each graph represent the corresponding continuous temporal evolution of the velocity distributions between dimensionless time 0 and 5. The distributions were computed from $10^5$ trajectories obtained by integrating the corresponding equations with a second-order Runge-Kutta method for stochastic differential equations (31). The values of the parameters are $\gamma\sqrt{I / kT} = 0.3$ and $(pE / kT) = 3$.

**Figure 3. Local heating and cooling without global temperature changes for an ensemble of dipoles.** The rotational kinetic temperature, $T_K = I\left\langle \omega^2 \right\rangle / k$, was computed as a function of time after switching off the field from the distributions of Fig. 2 for the whole system, $\theta \in \left(0, 2\pi\right)$, and for the regions of the angular coordinate with $\cos(\theta) < 0$ and $\cos(\theta) > 0$.

**Figure 4. Autonomous sorting of the equilibrium translational kinetic energy into nonequilibrated degrees of freedom upon switching off the trap for a gas of particles inside a box.** In terms of the dimensionless variables $\tau = t\big/\sqrt{m / K}$, $\chi = x\big/\sqrt{kT / K}$, and $\nu = v\big/\sqrt{kT / m}$, the dynamics is given by $d\chi / d\tau = \nu$ and $d\nu / d\tau = -\gamma\sqrt{m / K}\nu - \Theta(\tau)\chi + \infty(\delta(\chi + a') - \delta(\chi - a')) + \Gamma(\tau)$ with $\left\langle \Gamma(\tau)\Gamma(\tau') \right\rangle = 2\gamma\sqrt{m / K}\delta(\tau - \tau')$ and $a' = a\big/\sqrt{kT / K}$. The translational velocity distributions are shown at different times after switching off the trap for (**a**) the whole system, $a' > \left|\chi\right|$, and for the regions of the translational coordinate with (**b**)

$a' > |\chi| > \dfrac{a'}{2}$ and (**c**) $\dfrac{a'}{2} > |\chi|$ . The surfaces on the top-left of each graph represent the corresponding continuous temporal evolution of the velocity distributions between dimensionless time 0 and 5. The distributions were computed from $10^5$ trajectories obtained by integrating the corresponding equations with a second-order Runge-Kutta method for stochastic differential equations (31). The values of the parameters are $\gamma \sqrt{m / K} = 0.3$ and $a' = 4$ .

**Figure 5. Local heating and cooling without global temperature changes for a gas of trapped particles inside a box.** The translational kinetic temperature, $T_K = m \left\langle v^2 \right\rangle / k$ , was computed as a function of time after switching off the trap from the distributions of Fig. 4 for the whole system, $4 > \left| x \middle/ \sqrt{kT / K} \right|$ , and for the regions of the translational coordinate with $4 > \left| x \middle/ \sqrt{kT / K} \right| > 2$ and $2 > \left| x \middle/ \sqrt{kT / K} \right|$ .

**Figure 6. Breaking-off the ghost equilibrium requirements leads to far-from-equilibrium processes without thermal equilibrium-like velocity distributions.** Potential energy (**a**) and velocity distributions (**b**) are shown for the gas of trapped particles inside a box at different times after switching off a trap that is placed next to the right wall. The dynamics of the system is given by the same equations and parameter values as in Fig. 4 except for the force of the trap that is $-\Theta(\tau)(\chi - a')$ instead of $-\Theta(\tau)\chi$ . The value of the corresponding dimensionless time, $\tau = t \middle/ \sqrt{m / K}$ , is shown for each velocity distribution.

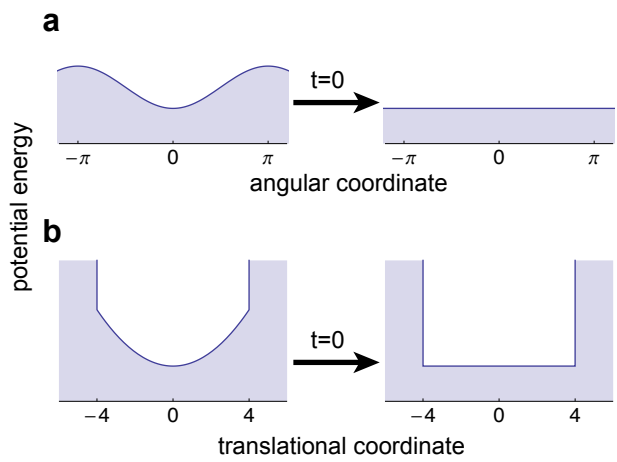

Figure 1

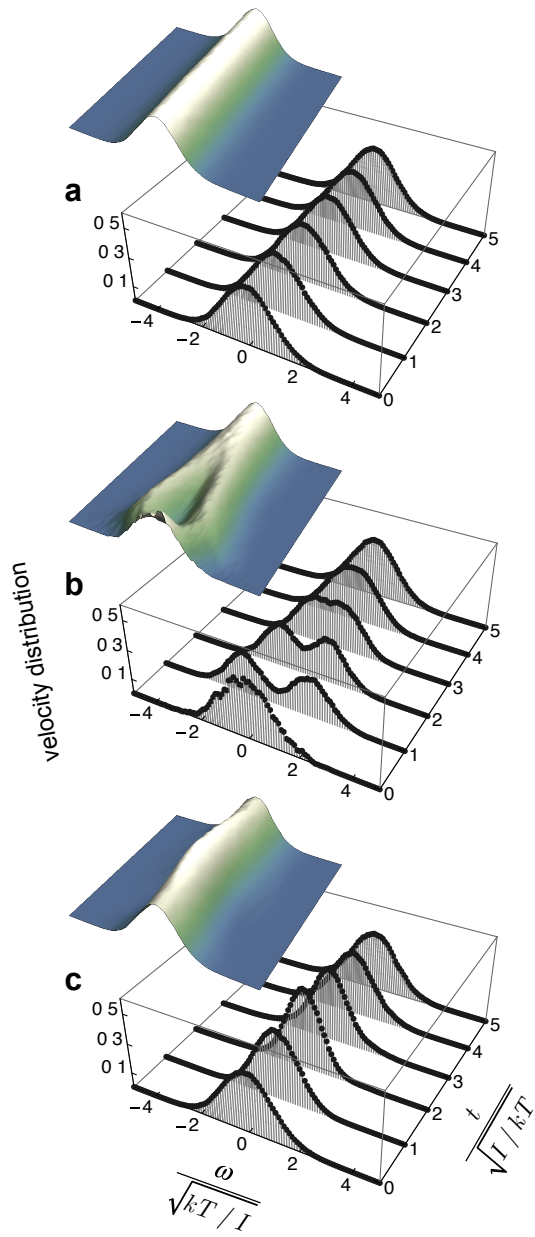

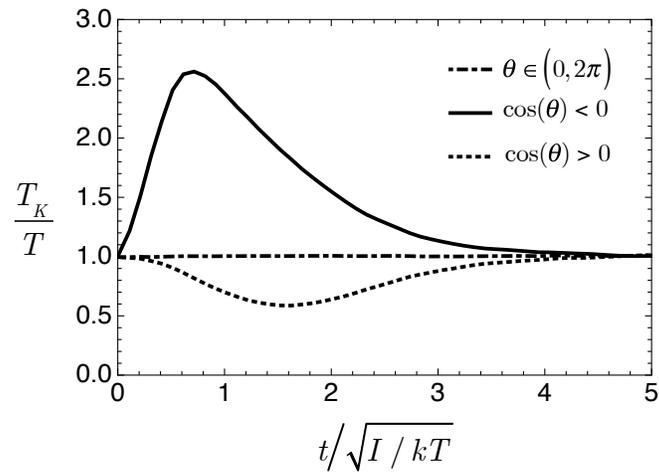

Figure 3

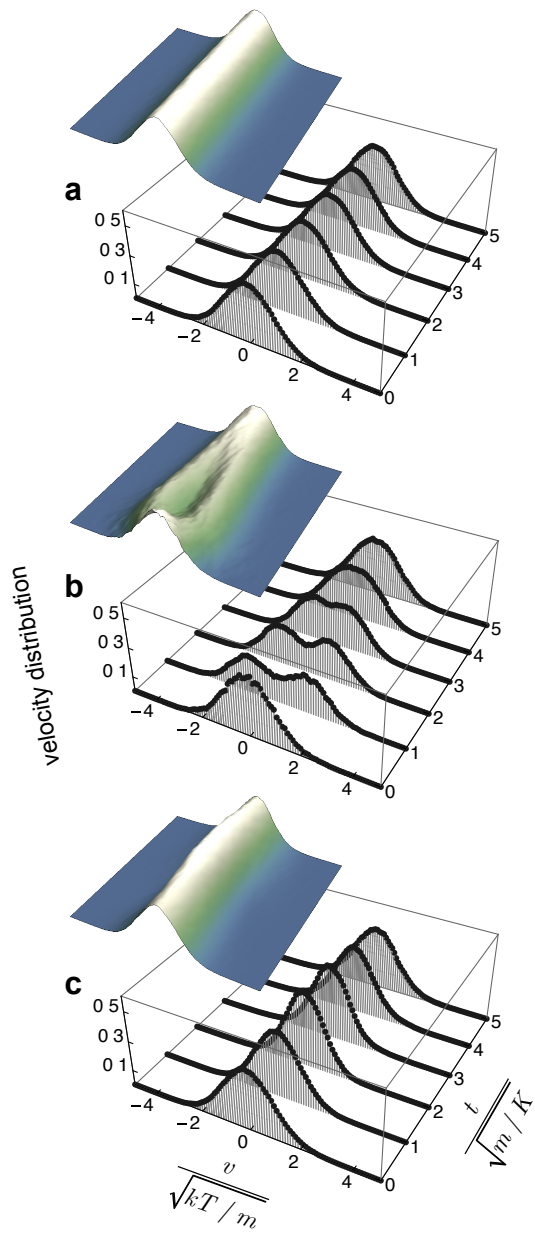

velocity distribution

**a**

**b**

**c**

$\dfrac{v}{\sqrt{kT/m}}$

$\dfrac{t}{\sqrt{m/K}}$

Figure 4

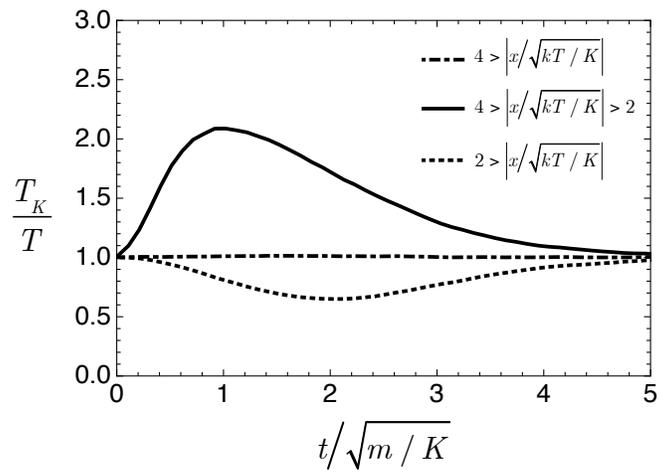

Figure 5

**a**

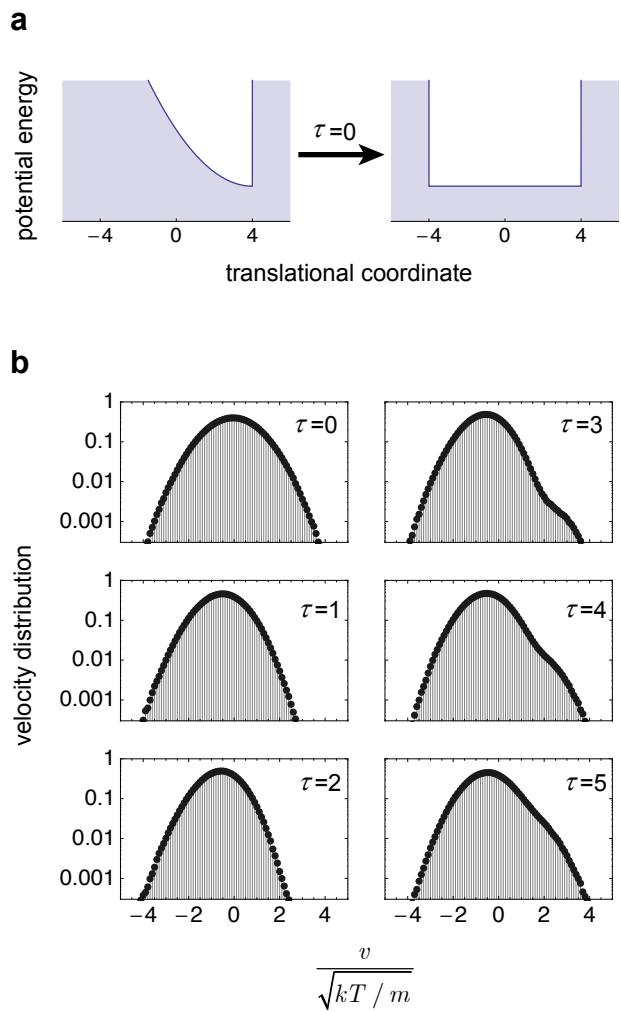

**b**

Figure 6